# Experimental Test of Analytic Theory of Spin Torque Oscillator Dynamics

C. Boone[1], J. A. Katine[2], J. R. Childress[2], J. Zhu[1], X. Cheng[1], I. N. Krivorotov[1]

1. *Department of Physics and Astronomy, University of California, Irvine, CA 92697*
2. *Hitachi Global Storage Technologies, San Jose, CA 95135*

## Abstract

We make measurements of power spectral density of the microwave voltage emitted by a spin torque nano-oscillator (STNO) and compare our experimental results to predictions of an analytic theory of a single-mode STNO dynamics [1]. We find that a complete set of the oscillator spectral properties: power, frequency, spectral line width and line shape as functions of current are self-consistently described by the analytic theory for moderate amplitudes of oscillations ($\lesssim 70°$).

Spin-polarized electric current injected into a metallic ferromagnet exerts a torque on its magnetization [2-4]. This spin transfer torque (STT) can act as negative magnetic damping and thereby excite persistent precession of magnetization [5]. Current-driven auto-oscillatory dynamics are usually observed in nanoscale spin valves [6-13] and nanocontacts to magnetic multilayers [14-17]. In these nanostructures, magnetization precession induced by a direct current gives rise to a microwave voltage signal. The power spectrum of this signal carries information on the STNO dynamics such as the oscillation frequency, amplitude and phase coherence [18,19]. Voltage-controlled microwave oscillator devices based on STNO demand high output power and narrow spectral line width, and understanding of the factors determining the spectral properties of the STNO signal is crucial for applications [20].

In this Letter, we report measurements of the power spectral density (PSD) of a STNO based on a nanocontact to a ferromagnetic nanowire. Our measurements reveal that the spectral line shape of this STNO is non-Lorentzian, and that the dependence of the STNO line width on current is surprisingly weak. To understand these findings, we compare our experimental results to predictions of a recently developed analytic theory of STNO [1]. We make a comparison of the *complete set* of the experimentally determined STNO spectral properties: power, frequency, line width and line shape as functions of current to predictions of the analytic theory. We find that within the range of its applicability, the analytic theory [1] self-consistently describes the dependence of all these spectral properties on current.

The free layer of the STNO studied in this Letter is a 6-nm thick, 50-nm wide permalloy (Py≡$Ni_{86}Fe_{14}$) nanowire on a Cu/Ta multilayer serving as the bottom electric lead. Spin-polarized current is injected into the nanowire through a 50×50 $nm^2$ $Co_{50}Fe_{50}$ (15 nm)/ Cu(8 nm) bilayer patterned on top of the Py nanowire as shown in Fig. 1(a), 1(b). The top Au/Ta lead is connected to the $Co_{50}Fe_{50}$ nanomagnet. This device is made in a multi-step nanofabrication process from a Ta(5)/ Cu(30)/ Ta(3)/ Cu(30)/ Ta(5)/ Cu(3)/ Py(6)/ Cu(8)/ $Co_{50}Fe_{50}$(15) / Cu(10)/ Ru(5)/ Ta(2.5) multilayer deposited by magnetron sputtering (thicknesses in parentheses are in nanometers). Saturation magnetization of the Py film ($M_S$=580 emu/$cm^3$) was measured by vibrating sample magnetometry. This reduced value of $M_S$ is consistent with 0.5-nm-thick "magnetically dead" layers at each Py/Cu interface. Figure 1(c) shows resistance of the sample as a function of magnetic field applied parallel to the nanowire axis. Micromagnetic simulations [21] of magnetoresistance for this device are in good agreement with the data in Fig. 1(c).

Spin torque dynamics in the Py nanowire are excited by applying a direct current, $I_{dc}$, between the top and bottom leads of the device with electrons moving from Py to $Co_{50}Fe_{50}$. Microwave voltage generated by the device is amplified and measured using a microwave spectrum analyzer [5,14]. For $I_{dc}$>1.95 mA, peaks at the first and second harmonics of the free layer precession frequency appear above the background noise in the microwave spectrum generated by the sample. We align the direction of the in-plane 500 Oe magnetic field with the nanowire axis by minimizing the integrated power in the first harmonic, $P_f$, relative to the power in the second harmonic, $P_{2f}$. From the $P_f/P_{2f}$ ratio, we estimate the misalignment angle between the magnetic field and the nanowire axis to be ~ 1° [5]. We studied eight STNO devices and observed similar PSD for all of them. All spectra reported in this Letter are measured at a bath temperature of 4.2 K, have zero-current background subtracted and are corrected for frequency-dependent attenuation and amplification in the microwave circuit used in the measurements.

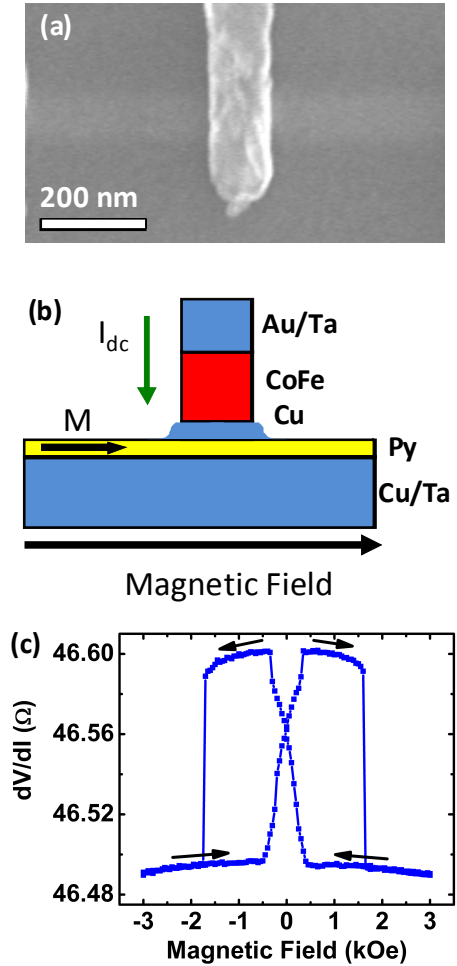

Figure 1 (color online): (a) SEM image of the nanowire (horizontal strip, faint due to a layer of alumina on top) and top lead (vertical strip). (b) Schematic of the device, which consists of a Py nanowire, a Cu spacer and a $Co_{50}Fe_{50}$ nanopillar with Au/Ta top lead. (c) Resistance versus magnetic field applied parallel to the Py nanowire, showing the $Co_{50}Fe_{50}$ nanomagnet switching at ±1.75 kOe and Py reversal in a ±350 Oe field range.

Figure 2(a) shows a typical PSD of the signal emitted by the device, and Fig. 2(b) summarizes the power spectra in the 1.8 – 3.5 mA current range. A single mode of oscillations is excited in the current intervals from 1.9 mA to 2.7 mA (mode 1) and from 3.0 mA to 3.4 mA (mode 2). In the current range from 2.7 mA to 3.0 mA both modes are present in the spectrum [22,23]. Micromagnetic simulations [21] show that magnetization precession in the Py nanowire is localized under the nanocontact due to the strong dipolar field from the $Co_{50}Fe_{50}$ nanomagnet. At $I_{dc} >$ 3.4 mA, no microwave signal is emitted by the sample. Magneto-resistance measurements show that a reverse domain is formed in the Py nanowire under the $Co_{50}Fe_{50}$ nanomagnet for $I_{dc} > 3.4$ mA.

To compare our data to an analytic theory of STNO [1], we convert the experimentally measured integrated power into a dimensionless power, $p$:

$$p = \sin^2\left(\frac{\theta_{max}}{2}\right) \quad (1)$$

where $\theta_{max}$ is the amplitude of in-plane oscillations of the free layer magnetization ($\theta(t) = \theta_{max} \cos(\omega t)$). To

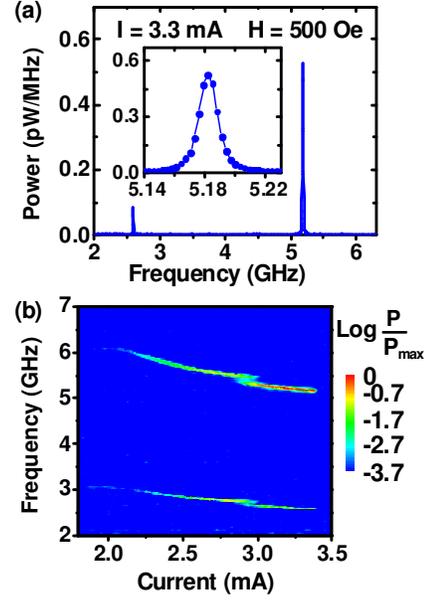

Figure 2 (color online): (a) Power spectrum of STNO at $I_{dc}$ =3.3 mA and $H$ = 500 Oe. Inset is a blow-up of the second harmonic. (b) Power spectrum at $H$= 500 Oe over the entire range of currents used in the measurements. The power abruptly disappears above 3.4 mA due to a reverse domain formation in the Py nanowire.

obtain $p$ from the experimental PSD, we assume the dependence of resistance on the angle between magnetizations of the Py and $Co_{50}Fe_{50}$ layers, $\theta$, given by:

$$R = R_0 + \Delta R_{max} \frac{\sin^2(\theta/2)}{1+\chi\cos^2(\theta/2)} \quad (2)$$

where $\Delta R_{max}$ = 0.105 $\Omega$ and $\chi$ is a magnetoresistance asymmetry parameter [23-25]. We have numerically verified that for $\theta_{max}$< 90°, $p$ is well described by:

$$p \approx \frac{1+\chi}{\chi + \frac{I_{dc}\Delta R_{max}}{4\sqrt{2R_s P_{2f}}}} \quad (3)$$

where $P_{2f}$ is the integrated power delivered to an impedance-matched load and $R_s$ = 46.55 $\Omega$ is resistance of the sample. The value of $\chi$ = 1.0±0.1 is determined from fitting the data of resistance versus magnetic field applied perpendicular to the sample plane. Figure 3(a) shows $p$ as a function of current calculated from Eq. (3) using experimentally measured $P_{2f}$ and $\chi = 1$. The values of $p$ in Fig. 3(a) indicate that magnetization precession with very large amplitude is excited. The minimum detectable precession amplitude is $\theta_{max} \approx 40°$ at $I_{dc} = 1.95$ mA while the maximum amplitude of $\theta_{max} = 88°$ is achieved at $I_{dc} = 3.375$ mA.

In the analytic theory of STNO [1], a stable precessional mode is achieved at $p$ for which negative damping due to STT, $\Gamma_-(p)$, is balanced by positive magnetic damping, $\Gamma_+(p)$. We assume a Slonczewski form of STT [26]:

$$\Gamma_-(p) = \sigma I_{dc} \frac{\Lambda^2(1-p)}{\Lambda^2 + (1-\Lambda^2)p} \qquad (4)$$

The nonlinear positive damping expanded to first order in $p$ is [1]:

$$\Gamma_+(p) \approx \Gamma_G(1+Qp) \qquad (5)$$

where $\sigma = \dfrac{\varepsilon \mu_B}{e \lambda \gamma}$, $\lambda = \dfrac{M_S V_0}{\gamma}$, $\Gamma_G = \dfrac{\alpha_G \omega_M}{2}$, $\omega_M = \gamma 4\pi M_S$, $\alpha_G$ is the Gilbert damping parameter, $V_0 = 1.5 \times 10^{-17}$ cm$^3$ is the volume of the Py nanowire under the Co$_{50}$Fe$_{50}$ nanomagnet, $\mu_B$ is the Bohr magneton, $e$ is the electron charge, $\gamma$ is the electron gyromagnetic ratio, $\varepsilon \leq 1$ is the degree of current spin polarization, $Q$ characterizes non-linear damping [27] and $\Lambda = \sqrt{\chi+1}$. Our analysis of the data with $Q \neq 0$ gives $Q = -0.02$, and since $Qp \ll 1$, we make a simplifying approximation $Q=0$. With this approximation, the stationary value of $p$ derived from the condition $\Gamma_+(p) = \Gamma_-(p)$ is:

$$p = \left(1 + \frac{1}{(\zeta - 1)\Lambda^2}\right)^{-1} \qquad (6)$$

where $\zeta = \frac{I_{dc}}{I_0}$ is the dimensionless current and $I_0 = \frac{\Gamma_G}{\sigma}$ is the critical current for excitation of persistent oscillations. We fit Eq. (6) to the experimental data of $p(\zeta)$ in Fig. 3(a) with a single fitting parameter, $I_0$, and obtain $I_0 = 1.9$ mA. The fit is good for $\zeta < 1.3$ ($p < 0.36$, $\theta_{max} < 73°$) and deviates from the experimental data at higher $\zeta$. In Fig. 3(b) we plot the angular frequency of the excited mode versus $p$ and find a linear relation between frequency and power $\omega(p) = \omega_0 + Np$ with $N = -4.8$ rad/ns for $p < 0.36$.

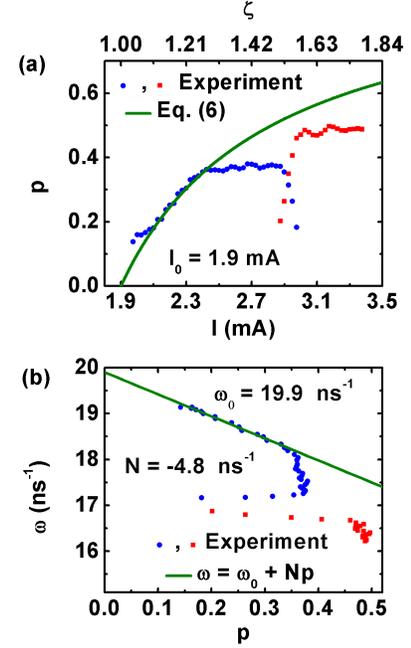

Figure 3 (color online): (a) Power versus applied current, showing the experimental data (symbols) and the best fit according to Eq. (6) (line) that gives the critical current $I_0 = 1.9$ mA. (b) Frequency versus power with a linear fit to $p < 0.36$ data.

Figure 3 demonstrates that the analytic theory of a single-mode STNO of Ref. [1] gives a good quantitative description of power and frequency as functions of current for moderate oscillation amplitudes ($\theta_{max} < 73°$). This theory also makes a prediction for the line shape and line width of the oscillator PSD, $S(\omega)$, using the same set of parameters that describe $p(\zeta)$ and $\omega(p)$. Therefore, a comparison of the experimentally measured line width and line shape to the theory's predictions can provide a comprehensive test of the theory. Specifically, the theory gives an expression for the oscillator autocorrelation function $\mathcal{K}(t)$ - the Fourier transform of the PSD, $S(\omega) = \int \mathcal{K}(t) \exp(i\omega t) dt$:

$$\mathcal{K}(t) = p \exp(i\phi(t)) \exp\left(-\Delta\omega_0 \left[\left(1+\left(\frac{Np}{\Gamma_p}\right)^2\right) t - \left(\frac{Np}{\Gamma_p}\right)^2 \frac{1-\exp(-2\Gamma_p t)}{2\Gamma_p}\right]\right) \quad (7)$$

Where $2\Delta\omega_0 = \Gamma_+(p)\frac{k_B T}{\lambda \omega(p) p}$ is the "linear" ($N=0$) oscillator line width, $k_B$ is the Boltzmann constant, $T$ is the STNO temperature, $\phi(t)$ is the oscillator phase and $\Gamma_p(p)$ is the relaxation rate of the oscillator towards its stationary power:

$$\Gamma_p(p) = p \frac{\partial}{\partial p'}(\Gamma_+(p') - \Gamma_-(p'))|_p \qquad (8)$$

Eq. (7) is valid for $p >> 2\pi\Delta f/|N|$, where $\Delta f$ ($2\Delta f$) is the full width at half maximum of the first (second) harmonic of the PSD.

The autocorrelation function given by Eq. (7) has a clear physical meaning. For $t << 1/\Gamma_p$, the STNO dephasing rate is $\Delta\omega_0$ - that of a linear oscillator ($N = 0$), while for $t >> 1/\Gamma_p$, the dephasing rate increases to $\Delta\omega_0\left(1+\left(\frac{Np}{\Gamma_p}\right)^2\right)$. Therefore, $\tau = 1/\Gamma_p$ is the delay time after which strong dephasing due to the STNO nonlinearity becomes dominant. We note that the characteristic dephasing time ~ $1/\Delta f$, of the STNO studied in this Letter, is similar to $\tau$. This means that neither of the two limits described above can be applied for a quantitative description of the STNO dephasing, and a full analysis based on Eq. (7) is required.

We obtain the experimentally determined autocorrelation function $\mathcal{K}(t)$ in Fig. 4(a) by a fast Fourier transform of $S_{ex}(2\omega)$, where $S_{ex}(\omega)$ is the measured PSD. This procedure is valid for STNO with $P_{2f} >> P_f$. For a quantitative analysis of the line width, we use $T \approx C \cdot I_{dc}$ [23]. We fit the function $\mathcal{K}(t) + const$, with $\mathcal{K}(t)$ given by Eq. (7) and the constant due to residual white noise, to the experimentally determined autocorrelation function with $\Gamma_G \cdot C$ and $\Gamma_p$ as fitting parameters. In the fitting procedure, we assume that $\Gamma_G \cdot C$ is $p$-independent while $\Gamma_p(p)$ varies with $p$. We find that the fit of $\mathcal{K}(t)$ is good for all current values

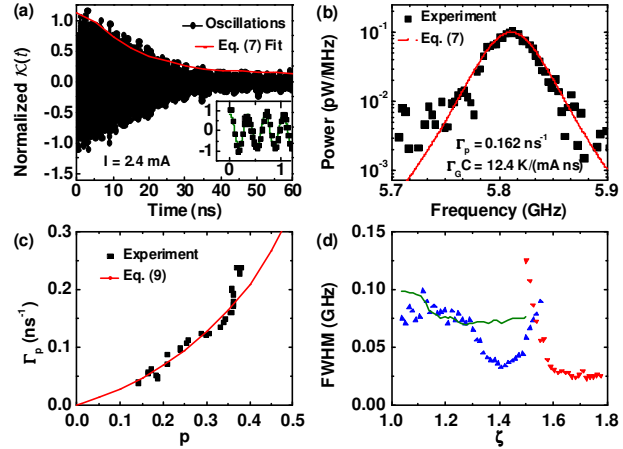

Figure 4 (color online): (a) Autocorrelation function $\mathcal{K}(t)$ of the STNO PSD with an envelope fit according to Eq. (7). Inset is a blow-up of $\mathcal{K}(t)$ in the $0 < t <1$ ns interval. (b) The second-harmonic spectral peak with Eq. (7) fit. (c) $\Gamma_p(p)$ obtained from $\mathcal{K}(t)$ fitting (symbols) and given by Eq. (9) (line). (d) Full widths at half maximum of the PSD second harmonic, $2\Delta f$, obtained from the experimental spectra (symbols) and given by theory according to Eq. (7) and Eq. (9) (line).

Page 7

up to the mode hopping regime ($I_{dc} < 2.7$ mA). The solid curve in Fig. 4(a) shows the best fit envelope function of $\mathcal{K}(t)$ for $I_{dc} = 2.4$ mA. This fitting procedure gives $\Gamma_G \cdot C = 12.4$ K/(mA·ns) and $\Gamma_p(p)$ shown in Fig. 4(c). The inverse Fourier transform of the best fit function $\mathcal{K}(t)$ is shown in Fig. 4(b) along with the experimental spectrum. The line shape given by Fourier transform of Eq. (7) is non-Lorentzian and describes the experimental data well.

The expected STNO relaxation rate $\Gamma_p(p)$ [1] is given by Eq. (8) and thus can be calculated using Eq. (4) and Eq. (6):

$$\Gamma_p(p) = \Gamma_G \frac{(1-p)\Lambda^2 + p}{(1-p)(\Lambda^2 + (1-\Lambda^2)p)^2} p \qquad (9)$$

The solid line in Fig. 4(c) shows the fit of Eq. (9) to the experimentally determined $\Gamma_p(p)$ with $\Gamma_G = \frac{\alpha_G \omega_M}{2}$ as a single fitting parameter. The value of $\alpha_G = 0.008$ obtained from this fit is similar to the Gilbert damping of Py films ($\alpha_G = 0.008$ [28]) and Py nanomagnets ($\alpha_G = 0.01$ [29]). This value of $\alpha_G$ corresponds to $C = 25$ K/mA similar to $C \approx 15$ K/mA determined from the measurements of resistance versus current and temperature [23]. Using the theoretical expressions of Eq. (9) and Eq. (7) and the experimental values of $p$ shown in Fig. 3(a), we calculate $2\Delta f$ and plot it in Fig. 4(d) as a function of current along with the measured $2\Delta f$. The measured and the theoretically predicted $\Gamma_p(p)$ and $2\Delta f$ are in a good agreement for $p < 0.36$ ($\zeta < 1.3$). The nonlinear theory predicts a decrease of the line width $\Delta f$ by a factor of 1.3 in the dimensionless power range 0.14 to 0.36 while the linewidth of a linear auto-oscillator is expected to decrease by factor of 2.6 ($\Delta \omega \sim 1/p$) [19]. Therefore, the non-linear theory of STNO successfully describes the weak variation of the line width with power and current observed in our experiment. Fig. 4(d) explicitly shows that the nonlinear theory of STNO [1] makes quantitative predictions of a single-mode STNO line width as a function of current given the experimental dependence of power on current. In the 1.3 < $\zeta$ < 1.5 current range, deviations of $p(\zeta)$ and $\Delta f(\zeta)$ from Eq. (6) and Eq.(9) are observed. According to Eq. (8), the experimentally observed sharp rise of $\Gamma_p(p)$ at $p = 0.36$ is consistent with the quasi-plateau in the $p(\zeta)$ at this power (Fig. 3(a)). Within the single-mode theory [1] the rise in $\Gamma_p(p)$, the quasi-plateau in $p(\zeta)$ and the decrease of $\Delta f$ (Fig. 4(d)) may be all explained by a deviation of the angular dependence of STT from that of Eq. (4) [30,25]. For $\zeta >1.5$, the oscillator enters a mode hopping regime and the single-mode theory is not applicable.

In conclusion, we compare the complete set of experimentally determined spectral properties of a STNO: power, frequency, spectral line width and line shape as functions of current to predictions of the analytic theory of STNO dynamics [1]. We find good agreement between the experimental data and the theory for the oscillator precession amplitudes below 70°. The analytic theory [1] successfully explains the weak dependence of the STNO line width on current and the non-Lorentzian spectral line shape. Our results suggest that measurements of the spectral shape of a STNO may serve as a tool for studies of the angular dependence of damping and spin transfer torque. This work is supported by the NSF Grants DMR-0748810 and ECCS-0701458, and by the Nanoelectronics Research Initiative through the Western Institute of Nanoelectronics.